# Remembrance: The Unbearable Sentience of Being Digital

Ragib Hasan
Department of Computer Science
University of Illinois at Urbana-Champaign
Urbana, IL 61801
rhasan@cs.uiuc.edu

Radu Sion
Department of Computer Science
Stony Brook University
Stony Brook, NY 11794
sion@cs.stonybrook.edu

Marianne Winslett
Department of Computer Science
University of Illinois at Urbana-Champaign
Urbana, IL 61801
winslett@cs.uiuc.edu

## ABSTRACT

We introduce a world vision in which data is endowed with memory. In this data-centric systems paradigm, data items can be enabled to retain all or some of their previous values. We call this ability *remembrance* and posit that it empowers significant leaps in the security, availability, and general operational dimensions of systems. With the explosion in cheap, fast memories and storage, large-scale remembrance will soon become practical. Here, we introduce and explore the advantages of such a paradigm and the challenges in making it a reality.

## 1. INTRODUCTION

Since the dawn of computing, data architectures have been essentially single-valued: each object instance is associated with only one value, namely, the value most recently assigned to it. Upon the next assignment, the old value of the object is overwritten. When one is given a digital object, its history (i.e., its past values and the actions that caused those values to change) is not usually available. Further, objects are usually not self-aware.

Many subcommunities of computer science have explored the idea that it would be beneficial to retain the history of data items, or at least their old values. For example, within the database community, we have seen version-based concurrency control [46], support for rollback to checkpoints [19, 44], the total recall introduced in Postgres [43, 42] that has evolved into time-travel databases [28, 35, 37] and temporal SQL [29, 39], and provenance for scientific data and workflows [3, 12, 13, 14, 36]. Outside the database community, there are proposed or actual systems for storing and recalling past system configurations [49], old versions of source code [5, 10], archival data [31], file system backups, old versions of individual files [26, 30, 32, 11, 40], old program and systems execution points in the form of checkpoints [33, 44] and recent history of an execution [41], past states of the World Wide Web on the WayBack machine [1], YouTube, and gMail-style email repositories, and more [2]. One can also think of logs as a poor-man's version of remembrance, and logs are heavily used in database systems, file systems [34], telecommunications and other critical infrastructures, and all sorts of applications. Though some of these approaches to remembrance keep old data because it is faster or simpler to keep it than to discard or overwrite it (e.g., log-structured file systems and gMail users), other approaches have found novel uses for the historical information. For example, many program crashes are due to transient failures, and many of these crashes can be avoided by retrying the last few instructions that were executed before the crash [41]. The US government has decided that Enron-style corporate fraud is such a threat to society's confidence in corporate America that companies need to keep a copy of every business email, spreadsheet, and report for several years [47], to enable after-the-fact prosecution of corporate evildoers. The proliferation of loyalty cards for retail shopping shows the benefits that can be obtained by mining a log of all shopping transactions. And the cultural anthropologists of the future will have a field day with the information obtainable from the WayBack machine [1].

Computer scientists have also proposed to endow objects with self-awareness, most notably in the programming language community [9, 22, 23, 38, 45, 48], but also much closer to home: for over 30 years we have been fond of associating each data item with its metadata, and have even toyed with the idea of stronger notions of self-awareness [27].

However, even at their grandest, most of these remembrance approaches boil down to *versioning systems* that ensure the persistence of explicitly defined versions of data objects. Because each approach was developed separately and for a different purpose, the result is a piecemeal coverage of past states of the world, with disparate, unintegrated interfaces reflecting a pastiche of different underlying assumptions. As a result, we cannot just click a button to go back to the state of the world as of 2 PM this day a year ago. Further, except for checkpointing, backup, data archival, and source code management, these versioning systems have not made it into the mainstream. Even where remembrance-based approaches *have* made it into the mainstream, a data item and its previous versions are not considered as an inseparable unit. Instead, each specific version is instantiated as a separate data item with its own unique identity. Associations between data items and their previous versions are maintained externally with great effort. These decisions were made for what were, at the time, good reasons. We believe that it is time to revisit those decisions in the light of recent changes in technology, and examine what benefits could accrue from having universal support for remembrance. It is also time to see what research contributions from the database community and other areas of computer science are needed in order to support universal integrated remembrance.

When thinking about remembrance in computing systems, it may be helpful to draw an analogy with the evolution of wetware [20], i.e., living beings. Single-celled organisms have little memory beyond what is hard-coded into their DNA. Remembering little, they can learn very little. Without much ability to learn, their potential to adapt dynamically to their environment is limited, and much of the pressure for adaptation to their environment is thrust onto reproductive processes that involve random changes in their DNA. As organisms became more complex, their memory and their ability to learn increased, as did their self-awareness. Sentient beings' memories retain vast amounts of contextual information with significant temporal components, often as explicit as the history of their formative elements. This meta-information has been shown to be an essential building block of humans' ability to reason, at the core of our associative memory processes [21]. This outlines one of the main differences between knowledge and raw information.

Although advanced wetware organisms such as humans can remember many events from their past history, they certainly cannot remember everything. For example, young humans have trouble remembering many things that their parents consider important, while their parents have trouble remembering where they left their car keys, and their grandparents have trouble remembering where they parked their car at the mall. At the dawn of the computer age, researchers were already positing that humans can benefit from computer assistance in remembering more of what happens to them [6]. Sixty years later, technology had advanced to the point that researchers could investigate so-called "total recall" paradigms for human activities such as the MyLifeBits [4, 16, 17, 18, 15] project, where both online and offline human activities are recorded in a database. These approaches use technology to provide a partial record (e.g., video and audio) of what happens to a person. "Total recall" is a misnomer for these experiments, which just remember what the human's recording device saw or heard, not how the human associated with the device felt about what was happening or how the human experienced the environment (e.g., hot, cold, sharp, soft, tense, relaxed, etc.). Thus it may be more accurate to classify these approaches as providing total recall not for a human, but rather for a device they carried around with them. However, even at the device level, such infrastructures currently lack self-awareness; for example, we cannot distinguish between the case where the video camera decided to turn itself off because its battery was low, and the case where the human switched off the camera. This lack of self-awareness is not surprising, since these projects are intended to endow humans with total recall, not humans' digital devices and their fine-grained data and application states. In the absence of system-level remembrance, only a partial temporal view of the life of a human or a digital artifact can be achieved. Ultimately, while Memex-inspired [6] technology such as MyLifeBits is for assisting humans, system-wide remembrance for data items can offer leaps in data processing capabilities.

We posit that, as computing evolves toward increasingly semantics-rich context-aware systems, a fundamentally novel data-with-memory model of remembrance will emerge. In this model, digital entities ranging from simple memory records to complex structured objects are inherently endowed with the ability to retain full or partial memories of their past contexts. Then history becomes an integral part of every entity. By considering historical values for data containers to be an integral part of the container, we enable processing based on not only the current incarnation of a data item, but also its past history, trends, evolution, and lineage. Data items stop being simple containers that live only in the current instant; they become aware of time, and have a knowledge of the evolution of their own state along the temporal axis. Such capabilities are essential to support the transition from information processing to knowledge processing.

In this paper, we define and explore this vision. While the long-term benefits' fruition is linked to the emergence of strong semantics processing paradigms, noteworthy advantages to deploying such a data model exist also in the immediate future. These range from significant leaps in the types and expressivity of the system policies now achievable, to self-healing systems that rely on remembrance to recover from undesirable events at the local level with minimal overall system impact, resulting in improved availability. To realize these benefits, however, we will have to overcome major technology hurdles in efficiently recording, storing, searching, indexing, retrieving, processing, and ignoring history. We must also seriously consider the possibility that intelligent behavior requires a well-developed ability to forget, as well as to remember; this raises the question of how we should choose what to forget.

## 2. WHAT TO REMEMBER?

At the extreme, we can imagine a system where every component, from the micro-level to the macro-level, remembers everything that has happened to it, from the moment of its creation through the moment of its destruction. At the lowest levels of abstraction, memory locations can have remembrance, making it possible to query for older values of data stored at a particular location. Individual data blocks in secondary storage can have memory of their previous context and values. Higher logical structures such as files are no longer dumb containers of data; they retain contextual information about their contents throughout their lifetime. More abstractly, a self-aware PDF file remembers the LaTex or Word documents from which it was created, even though they reside in separate containers. Variables in an execution of a program can possess remembrance, including awareness of the environmental conditions that affected their execution. The program itself can remember its executions and their effects, as can the larger configurations that include it. Inside a database, we may choose to make tuples, schemas, constraints, triggers, stored procedures, and other metadata remembrance-capable, as in an extension of time-travel databases. We can do the same for transactions and transaction executions.

Even application-layer constructs and data containers can have remembrance. A visitor to a web page can get not only the current instance of the web page, but also traverse the page chronologically. Web searches can include the temporal dimension; this is already possible now with Google News where users can specify a particular time frame when searching for old news. Queries no longer return a flat time-ignorant result, but rather a result that can be browsed along the temporal axis.

Transfer, copying, and movement of data objects can preserve memory – a data container can retain its memory when it is moved to a new location. The *copy* operation transfers old memory from the original source to the copied container, along with new memory of the copy operation. Deletion removes the container, but its memories can live on, as in time-travel databases.

Is this vision attainable, or even desirable? Storage is cheap, but in practice, physical, logistical, legal, and pragmatic issues will limit what a computer system can and should remember.

**Physical limits.** Hardware does fail and bits do rot, even on magnetic disk. When hardware dies, any memories stored on it will die with it, so memories that should not be lost should be stored in a fault-tolerant manner. In a system that supports remembrance, users will expect their data to last forever, with no exceptions for bit rot, human error, or hardware failure. Large-scale fault-tolerance for remembrance will probably be easiest to provide in a cloud computing context, such as in Google's initiative to preserve publicly sharable scientific data.

No hardware support is available for remembrance at the level of physical bytes, either in L2 cache, memory, or disk. Further, the astronomical cost of such support and the security issues that it would introduce outweigh the benefits that might result from, e.g., improved forensic analysis. Thus we restrict our attention to support for remembrance at higher levels of abstraction, such as abstract locations.

On the other hand, storage has become very cheap, with the result that we already retain much more data than we used to. The cost of accessing that data has dropped much more slowly, but that will not matter if historical information is rarely accessed.

**Performance limits.** At any level of abstraction, a naive implementation of remembrance will kill CPU, network, and system performance. For example, suppose that a fetch of an object from disk brings not only the current value of the object, but also all its previous values and its other associated memories. If the application does not need this extra information, it will occupy valuable buffer space and pollute the L2 cache, crippling performance. Thus historical information should be available when wanted and invisible the rest of the time. For example, we may prefer to keep historical versions of tuples on separate file pages from the latest versions. At an extreme, we might choose to bury the historical information in logs and only dredge it up on request. In other words, performance considerations may dictate that we develop a very clean user interface that provides the illusion of a self-aware and history-aware system. In the relatively rare event that historical information is accessed through this interface, we will quickly cobble it together from searchable, indexable logs that we have shipped off to inexpensive remote self-organizing storage. This extreme vision raises many new challenges in how to transparently move logs off to a cloud of inexpensive, fault-tolerant storage; how to transparently reorganize and/or index it in a manner conducive to future access patterns; and how to transparently decide where to place the data and its replicas. These problems are particularly acute for multimedia data, such as sounds, images, and videos. For example, suppose that we have a MyLifeBits record of our life. How quickly can we get an answer to the question *Where are my car keys?* or *Where did I park the car?* Is it a security violation for the system to answer the question *Where did my spouse park the car?* What about *Where did my teenager park the car?* or simply *Where is my teenager?* Imagine the benefit for a blind person if the system can answer questions of the form *Where did I leave my house keys/wallet/comb?* These examples show that both the utility and the sensitivity of a remembrance are very context-sensitive.

**Security and legal challenges.** As suggested by the examples above, some remembrances are very sensitive, and it will be hard to manage them in a user-friendly way that preserves confidentiality and privacy. (In the security community, *privacy* refers to the ability of the owner of a piece of information to control what is done with it.) For example, Microsoft Word's limited ability to remember the history of a document has already caused scandals where the recipient of a document was able to view previous drafts of the document, and publicized their contents. Imagine the complications if a PDF of a recommendation letter was inseparable from the versions of the Word file from which it was created! If we cannot erase sensitive remembrances from a disk, then the disk cannot be given to a new owner. Similarly, shared computing facilities rely on our ability to erase the memory of executions of sensitive programs. Digital rights management facilities often depend on computers' ability to forget digital information. Laws such as HIPAA [8] require that certain kinds of electronic records only be kept for a limited period of time, and be destroyed thereafter. Thus we need iron-clad ways to ensure that sensitive remembrances either do not fall into inappropriate hands, or are unintelligible if they do.

**Usability challenges.** Remembrances may be used directly by a self-aware computer system to tune and improve itself, or by a human who needs historical information. Both kinds of interfaces will be challenging to provide at an appropriate level of abstraction. This problem has been addressed in individual history-aware systems, but never in a manner intended to span multiple independent systems with autonomy.

**Philosophical challenges.** The preceding discussion raises fundamental questions about our current notions of what an object is and what object identity means.

**Limiting remembrance.** The foregoing discussion suggests that even if we could, it may be better for computer systems not to remember everything. The fact that living beings tend to forget things also suggests that there is value in the ability to forget. Psychological studies of humans who do remember everything [24, 7] strongly support this conclusion, explaining that people cursed with a perfect memory (a medical condition called Superior Autobiographical Memory Syndrome or *Hyperthymesia*) find themselves overwhelmed by memories, distracted by memories, and/or unable to abstract away from the details of their memories. Thus we postulate that systems with remembrance also need the ability to forget – immediately raising the question of what to forget.

In some situations, specific hard-coded policies will dictate what to forget. For example, companies often prefer to delete routine business documents once their mandated retention period has ended [25]. This policy is in place because companies have learned that the legal liabilities that result from retaining such documents, which can be subpoenaed in lawsuits, exceed their internal value to the company. This raises the larger question of how a computer system can learn such policies automatically.

An analogy to the short-term and long-term memory of humans may be helpful here. We retain small details in memory for a short period because they are most likely to be useful during that period. For example, a person (let us call her Alice) can remember what she ate for lunch today; that knowledge may guide her choice of food for dinner. If Alice tried hard, she could probably remember what she ate for lunch two days ago. But Alice will not be able to remember what she ate for lunch one month ago. Her memory has automatically removed that detail because she has not made use of it for an extended period. This suggests that as a default policy, routine remembrances in a computer system may fade away gradually, becoming less easily accessible over time, until they reach a threshold where they are forgotten entirely. For example, the details of the execution of a program may no longer be worth remembering once the program has finished executing. However, if the program is considered important, such as a financial transaction or a change in the system configuration, the remembrance may be retained for an extended period, in logs or in other forms. Further, there may be high value in remembering the recent details of the execution of a program, because they can allow us to recover automatically from certain classes of failures [41].

In humans, detailed memories are often replaced by general patterns that we learn from them. For example, we learn that Valentine's Day often involves giving and receiving flowers, cards, and candy, though we may forget the details of individual instances of Valentine's Day. Similarly, before remembrances fade, the computer system should use data mining techniques to learn whatever useful patterns it can glean from them. At the simplest level, such techniques can be used to improve the tuning of the system and plan future allocations of resources. Further, with system-wide remembrance in place, we will have new opportunities for automated tuning and learning.

## 3. CONCLUSION

We have briefly presented our vision of the opportunities and challenges of a world where computing systems are self-aware and remember important aspects of their history and evolution. While bits and pieces of this vision already exist in some applications, these pieces have never been tied together into a seamless continuum. If we can make systems sentient of their old context, lineage, and values, and address the resulting challenges for performance and usability, then we have the potential to reach new levels of self-tuning in computer systems and support many exciting new user-level applications.

## Acknowledgements

Hasan and Winslett were supported in part by NSF awards CNS-0716532 and CNS-0803280. Sion was supported in part by the NSF through awards CNS-0627554, CNS-0716608, CNS-0708025, and IIS-0803197. Sion would also like to thank Motorola Labs, Xerox PARC, IBM Research, the IBM Software Cryptography Group, CEWIT, and the Stony Brook Office of the Vice President for Research.## 4. REFERENCES

[1] Internet Archive. WayBack Machine. Online at http://www.archive.org.

[2] Michael Arrington. Swivel aims to become the internet archive for data. TechCrunch, Online at http://tinyurl.com/ydnrcc, December 2006.

[3] Roger S. Barga and Luciano A. Digiampietri. Automatic generation of workflow provenance. In Luc Moreau and Ian T. Foster, editors, *Proceedings of the International Provenance and Annotation Workshop*, volume 4145 of *Lecture Notes in Computer Science*, pages 1–9. Springer, 2006.

[4] Gordon Bell, Jim Gemmell, and Roger Lueder. Challenges in using lifetime personal information stores. In *Proceedings of the 27th Annual International ACM SIGIR conference on Research and Development in Information Retrieval*, New York, NY, USA, 2004. ACM.

[5] Brian Berliner. CVS II: parallelizing software development. In *Proceedings of the Winter 1990 USENIX Conference*, pages 341–352. USENIX Assoc., 1990.

[6] Vannevar Bush. As we may think. *The Atlantic Monthly*, 176(1):101–108, 1945.

[7] Patricia Casey. Unravelling the mysterious curse of a flawless memory. The Independent, Online at http://tinyurl.com/4rwm8s, October 6 2008.

[8] Centers for Medicare & Medicaid Services. The Health Insurance Portability and Accountability Act of 1996 (HIPAA). Online at http://www.cms.hhs.gov/hipaa/, 1996.

[9] Pierre Cointe. Reflective languages and metalevel architectures. *ACM Comput. Surv.*, page 151, 1996.

[10] Ben Collins-Sussman. The subversion project: building a better CVS. *Linux Journal*, 2002(94), 2002.

[11] Brian Cornell, Peter A. Dinda, and Fabián E. Bustamante. Wayback: a user-level versioning file system for Linux. In *Proceedings of the USENIX Annual Technical Conference*, pages 19–28, Berkeley, CA, USA, 2004. USENIX Association.

[12] Yingwei Cui, Jennifer Widom, and Janet L. Wiener. Tracing the lineage of view data in a warehousing environment. *ACM Trans. Database Syst.*, 25(2):179–227, 2000.

[13] Ian T. Foster, Jens-S. Vockler, Michael Wilde, and Yong Zhao. Chimera: A virtual data system for representing, querying, and automating data derivation. In *Proceedings of the 14th International Conference on Scientific and Statistical Database Management*, pages 37–46, Washington, DC, USA, 2002. IEEE Computer Society.

[14] James Frew and Rajendra Bose. Earth system science workbench: A data management infrastructure for earth science products. In *Proceedings of the Thirteenth International Conference on Scientific and Statistical Database Management (SSDBM '01)*, page 180, Washington, DC, USA, 2001. IEEE Computer Society.

[15] Jim Gemmell, Gordon Bell, and Roger Lueder. MyLifeBits: a personal database for everything. *Commun. ACM*, 49(1):88–95, 2006.

[16] Jim Gemmell, Gordon Bell, Roger Lueder, Steven Drucker, and Curtis Wong. MyLifeBits: fulfilling the memex vision. In *Proceedings of the 10th ACM international Conference on Multimedia*, pages 235–238, New York, NY, USA, 2002. ACM.

[17] Jim Gemmell, Roger Lueder, and Gordon Bell. The MyLifeBits lifetime store. In *Proceedings of the ACM SIGMM Workshop on Experiential Telepresence*, pages 80–83, New York, NY, USA, 2003. ACM.

[18] Jim Gemmell, Lyndsay Williams, Ken Wood, Roger Lueder, and Gordon Bell. Passive capture and ensuing issues for a personal lifetime store. In *Proceedings of the 1st ACM Workshop on Continuous Archival and Retrieval of Personal Experiences*, pages 48–55, New York, NY, USA, 2004. ACM.

[19] Jim Gray and Andreas Reuter. *Transaction Processing: Concepts and Techniques*. Morgan Kaufmann Publishers Inc., San Francisco, CA, USA, 1992.

[20] Vaiva Kalnikaité and Steve Whittaker. Software or wetware?: discovering when and why people use digital prosthetic memory. In *Proceedings of the SIGCHI Conference on Human Factors in Computing Systems*, pages 71–80, New York, NY, USA, 2007. ACM.

[21] Eric R. Kandel, James H. Schwartz, and Thomas M. Jessell. *Principles of Neural Science*. McGraw-Hill Medical, 2000.

[22] Fabio Kon, Fabio Costa, Gordon Blair, and Roy H. Campbell. The case for reflective middleware. *Commun. ACM*, 45(6):33–38, 2002.

[23] Jeff Kramer and Jeff Magee. Self-managed systems: an architectural challenge. In *2007 Future of Software Engineering*, pages 259–268, Washington, DC, USA, 2007. IEEE Computer Society.

[24] Ariel Leve. Jill Price, the woman who remembers everything. Times Online, UK. Online at http://www.timesonline.co.uk/tol/news/uk/science/article4771978.ece, September 21 2008.

[25] Soumyadeb Mitra, Marianne Winslett, and Nikita Borisov. Deleting index entries from compliance storage. In *Proceedings of the 11th International Conference on Extending Database Technology*, pages 109–120, New York, NY, USA, 2008. ACM.

[26] Kiran-Kumar Muniswamy-Reddy, Charles P. Wright, Andrew Himmer, and Erez Zadok. A versatile and user-oriented versioning file system. In *Proceedings of the 3rd USENIX Conference on File and Storage Technologies*, pages 115–128, Berkeley, CA, USA, 2004. USENIX Association.

[27] Lars E. Olson, Carl A. Gunter, and P. Madhusudan. A formal framework for reflective database access control policies. In *Proceedings of the 15th ACM Conference on Computer and Communications Security (CCS '08)*, Alexandria, VA, October 2008.

[28] Michael A. Olson. The design and implementation of the inversion file system. In *USENIX Winter*, pages 205–218, 1993.

[29] Gultekin Ozsoyoglu and Richard Thomas Snodgrass. Temporal and real-time databases: A survey. *IEEE Trans. on Knowl. and Data Eng.*, 7(4):513–532, 1995.

[30] Zachary Peterson and Randal Burns. Ext3cow: a time-shifting file system for regulatory compliance. *Trans. Storage*, 1(2):190–212, 2005.

[31] Zachary N. J. Peterson and Randal Burns. Building regulatory compliant storage systems. In *Proceedings of the 2006 International Conference on Digital Government Research*, pages 442–443, New York, NY, USA, 2006. ACM.

[32] Zachary N. J. Peterson, Randal Burns, Giuseppe Ateniese, and Stephen Bono. Design and implementation of verifiable audit trails for a versioning file system. In *Proceedings of the 5th USENIX Conference on File and Storage Technologies (FAST '07)*, pages 93–106, Berkeley, CA, USA, 2007. USENIX Association.

[33] Feng Qin, Joseph Tucek, Yuanyuan Zhou, and Jagadeesan Sundaresan. Rx: Treating bugs as allergies—a safe method to survive software failures. *ACM Trans. Comput. Syst.*, 25(3):7, 2007.

[34] Mendel Rosenblum and John K. Ousterhout. The design and implementation of a log-structured file system. *ACM Trans. Comput. Syst.*, 10(1):26–52, 1992.

[35] Ross Shaull, Liuba Shrira, and Hao Xu. Skippy: a new snapshot indexing method for time travel in the storage manager. In *Proceedings of the 2008 ACM SIGMOD International Conference on Management of Data*, pages 637–648, New York, NY, USA, 2008. ACM.

[36] Yogesh L. Simmhan, Beth Plale, and Dennis Gannon. A survey of data provenance in e-science. *SIGMOD Rec.*, 34(3):31–36, September 2005.

[37] A. Prasad Sistla and Ouri Wolfson. Temporal conditions and integrity constraints in active database systems. In *Proceedings of the 1995 ACM SIGMOD International Conference on Management of Data*, pages 269–280, New York, NY, USA, 1995. ACM.

[38] Brian Cantwell Smith. Reflection and semantics in LISP. In *Proceedings of the 11th ACM SIGACT-SIGPLAN symposium on Principles of programming languages*, pages 23–35, New York, NY, USA, 1984. ACM.

[39] Richard T. Snodgrass, editor. *The TSQL2 Temporal Query Language*. Kluwer, 1995.

[40] Craig A. N. Soules, Garth R. Goodson, John D. Strunk, and Gregory R. Ganger. Metadata efficiency in versioning file systems. In *Proceedings of the 2nd USENIX Conference on File and Storage Technologies (FAST'03)*, pages 43–58, Berkeley, CA, USA, 2003. USENIX Association.

[41] Sudarshan M. Srinivasan, Srikanth Kandula, Christopher R. Andrews, and Yuanyuan Zhou. Flashback: a lightweight extension for rollback and deterministic replay for software debugging. In *Proceedings of the USENIX Annual Technical Conference*, pages 29–44, Berkeley, CA, USA, 2004. USENIX Association.

[42] Michael Stonebraker. The Postgres DBMS. In *SIGMOD Conference*, page 394, 1990.

[43] Michael Stonebraker and Greg Kemnitz. The Postgres next generation database management system. *Commun. ACM*, 34(10):78–92, 1991.

[44] Paula Ta-Shma, Guy Laden, Muli Ben-Yehuda, and Michael Factor. Virtual machine time travel using continuous data protection and checkpointing. *SIGOPS Oper. Syst. Rev.*, 42(1):127–134, 2008.

[45] Eno Thereska, Dushyanth Narayanan, and Gregory R. Ganger. Towards self-predicting systems: What if you could ask 'what-if'? *Knowl. Eng. Rev.*, 21(3):261–267, 2006.

[46] Robert H. Thomas. A majority consensus approach to concurrency control for multiple copy databases. *ACM Trans. Database Syst.*, 4(2):180–209, 1979.

[47] U.S. Public Law No. 107-204, 116 Stat. 745. The Public Company Accounting Reform and Investor Protection Act, 2002.

[48] Takuo Watanabe and Akinori Yonezawa. Reflection in an object-oriented concurrent language. In *Proceedings of the Conference on Object Oriented Programming Systems Languages and Applications*, pages 306–315. ACM New York, NY, USA, 1988.

[49] Andrew Whitaker, Richard S. Cox, and Steven D. Gribble. Configuration debugging as search: finding the needle in the haystack. In *Proceedings of the 6th Symposium on Opearting Systems Design & Implementation*, Berkeley, CA, USA, 2004. USENIX Association.